# Single crystal growth and properties of $MgB_2$ and $Mg(B_{1-x}C_x)_2$

S.M. Kazakov[a], J. Karpinski[a], J. Jun[a], P. Geiser[a], N.D. Zhigadlo[a], R. Puzniak[b], A.V. Mironov[c]

[a]*Solid State Physics Laboratory ETH, 8093-Zürich, Switzerland,* [b]*Institute of Physics, Polish Academy of Sciences, 02-668 Warsaw, Poland,* [c]*Department of Chemistry, Moscow State University, 119899 Moscow, Russia*

**Abstract**

Single crystals of $MgB_2$ and $Mg(B_{1-x}C_x)_2$ have been grown using cubic anvil technique. $T_c$ values vary in a wide range (39-9 K) with carbon content varying from 0 up to 16%. Using SiC as the precursor leads to C and not to Si substituted crystals. Micro-hardness measurements performed on $MgB_2$ single crystals give average value of 1100 kg/mm$^2$.

## 1. Introduction

In spite of the simple crystal structure and chemical composition of $MgB_2$, its single crystal growth is a very difficult task. The largest single crystals can be obtained only at high hydrostatic pressure [1,2]. High-pressure high-temperature technique prevents the evaporation of Mg and increases the solubility of $MgB_2$ in molten Mg. This increases the efficiency of the crystal growth process. Many efforts have been made to modify superconducting properties of this compound through chemical substitution. Carbon substitution for boron appeared to be one of the most interesting, though reports of the carbon solubility in $MgB_2$ and influence on $T_c$ varied considerably. Neutron diffraction study determined maximum carbon substitution level to be about 10% and proposed relation between *a* lattice parameter and C concentration [3]. Dou et al. reported an enhancement of the critical current density and flux pinning of $MgB_2$ powder by nanoparticle SiC doping [4]. They claimed that simultaneous Si and C substitution for B raised the saturation limit considerably (up to x=0.65) while $T_c$ reduction is not pronounced (only 2.4 K). High resolution X-ray diffraction investigations indicate presence of two $MgB_2$:C phases [5]. All these substitutional experiments have been performed so far on polycrystalline samples. However polycrystalline samples measurements give only average data of collection of crystallites. The investigations of single crystals could help to resolve the issue concerning solubility limit and $T_c$ variation as a function of doping. In this paper, we report the results of the single crystal growth of pure and carbon substituted $MgB_2$ as well as $T_c$ dependence on C content. In addition, we report for the first time the results of hardness measurements on $MgB_2$ crystals.

## 2. Crystal growth

Crystal growth experiments, using the cubic anvil system, have been performed at P=30 kbar. In first type of precursors, SiC was added to a mixture of Mg and amorphous B. In a second case, graphite served as a source of carbon. Starting materials with a different nominal carbon content were mixed, pressed and placed in BN crucibles. First, pressure of 30 kbar was applied using a pyrophylite cube as a medium, then the temperature was increased during one hour, up to the maximum of 1900-1950°C, kept for 30 min, and decreased during 1-2 hours.

## 3. Results

Carbon substituted $Mg(B_{1-x}C_x)_2$ crystals were grown with dimensions up to 0.8x0.8x0.02 $mm^3$. They were black in color in contrast to golden non-substituted $MgB_2$. EDX and laser ablation ICP mass spectroscopy analyses performed on the SiC-doped crystals show only C and no traces of Si in the crystals. X-ray single crystal analysis confirmed this finding. Carbon content of the crystals was estimated from *a* lattice parameter according to [3], assuming linear dependence of *a* parameter on carbon content. Magnetization curves of crystals with various carbon contents are shown in Fig.1. Figure 2 shows the variation of $T_c$ as a function of carbon concentration. $T_c$ dependence on C content is monotonic and different from this on Al content [1], which achieves discrete values. By adjusting carbon content one can tune $T_c$ in a wide range (39-9 K) for carbon content between x=0 and 0.16.

The decrease of lattice parameter *a* is quite pronounced (from 3.0844 Å for x=0 to 3.0344 Å for x=0.16) while *c*-parameter is almost constant, which is an agreement with other reports. Structural studies do not prove the existence of two phases or superstructures in carbon substituted crystals, although from reflection profiles and their orientation we may suppose the existence of some disorder in the substituted crystals.

## 3. Hardness of $MgB_2$ single crystals

Micro-hardness investigations of $MgB_2$ single crystals have been performed with the Knoop method. Load of 50 gramms was applied for 30 seconds per imprint. The force was applied along the *c*-axis of the crystals. The crystals were embedded in Technovit 4004. Crack formation and imprint asymmetry can be observed in some crystals, depending on the crystal morphology. For evaluation of the hardness, only crack free and symmetric imprints were evaluated. Seven crystals from the same batch were probed, yielding 10 datapoints from 28 attempts. The values range from 790 kg/mm$^2$ to 1288 kg/mm$^2$, with average value of 1100 kg/mm$^2$. Comparable values are found for measurements on freestanding crystals. For comparison, Knoop hardness of Quarz is 750 kg/mm$^2$ and this of Topaz is 1250 kg/mm$^2$. This proves, that $MgB_2$ is a hard material, hence crystals are britle.

**References**

1. J. Karpinski et al., this proceedings.
2. J. Karpinski et al. Physica C 385 (2003) 42.
3. M. Avdeev et al., Physica C 387 (2003) 301.
4. S.X. Dou et al., Appl. Phys. Lett. 81 (2002) 3419.
5. I. Maurin et al., Physica B 318 (2002) 392.

**Figure captions.**

Figure 1. Fig.1. Susceptibility measurements of C substituted $MgB_2$ crystals.

Figure 2. $T_c$ dependence on C content in $Mg(B_{1-x}C_x)_2$ crystals.

Figure 3. Knoop micro-hardness measurement on $MgB_2$ single crystal.

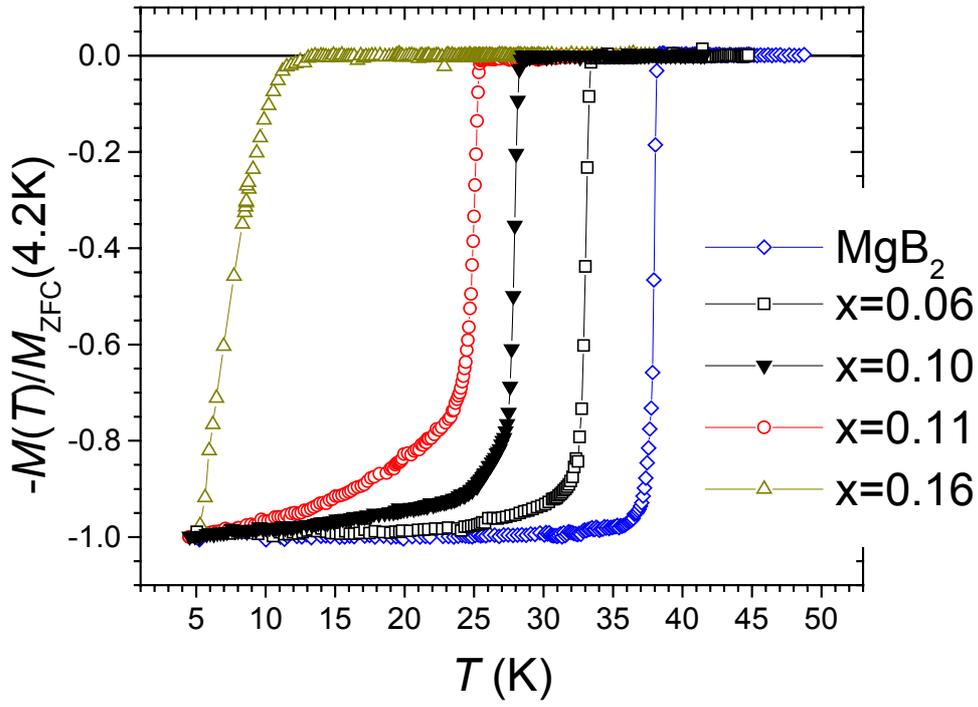

Figure 1.

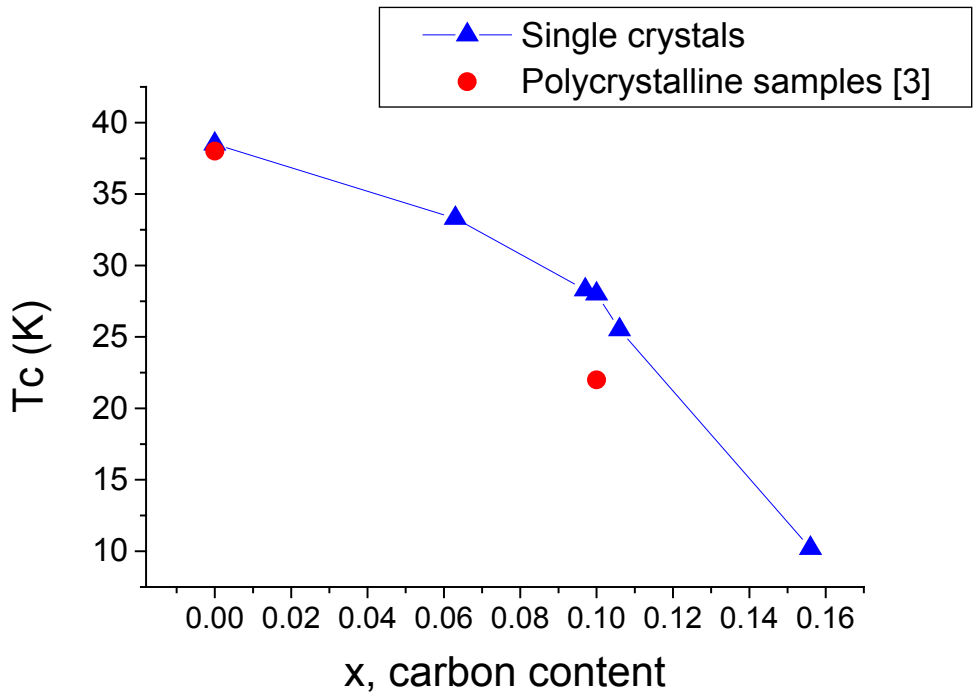

Figure 2

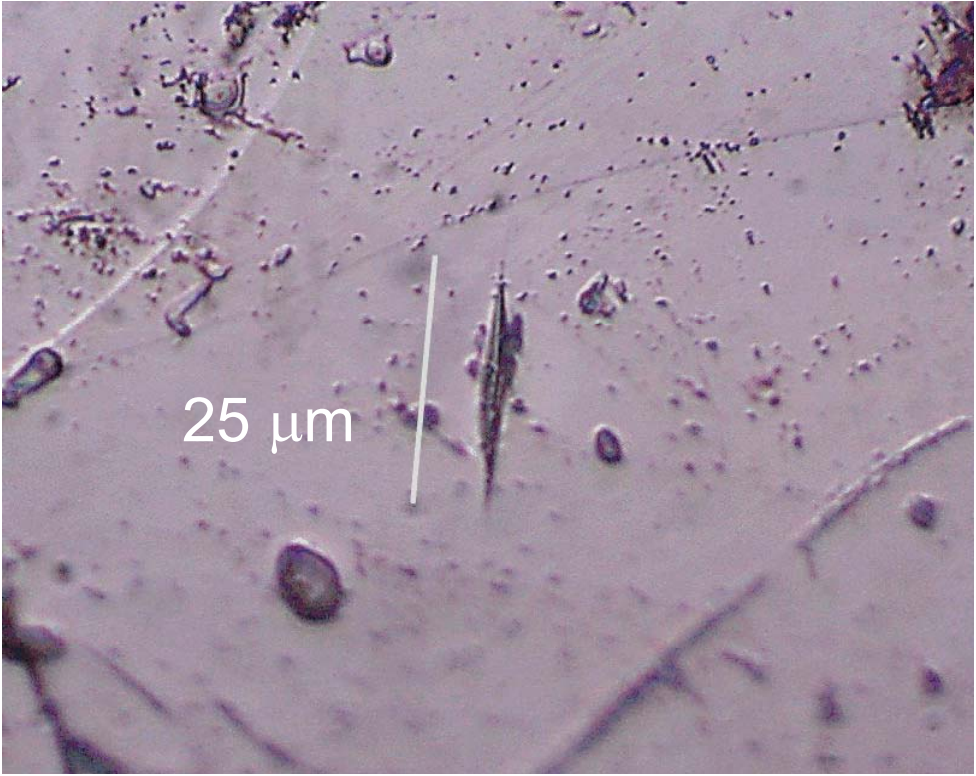

Figure 3.